\documentstyle[aps,multicol]{revtex}
\renewcommand{\narrowtext}{\begin{multicols}{2} \global\columnwidth20.5pc}
\renewcommand{\widetext}{\end{multicols} \global\columnwidth42.5pc}
\newcommand{\p}{\partial}
\newcommand{\D}{\Delta}
\newcommand{\e}{\varepsilon}

\begin{document}

\title{Universality of equilibrium one-dimensional transport
from gauge invariance}

\author{Anton Yu. Alekseev$^{* \dagger \ddagger}$,
 Vadim V. Cheianov$^{* \dagger \ddagger }$, 
J\"{u}rg Fr\"{o}hlich$^{\dagger}$}

\address{$^*$
Institutionen f\"{o}r Teoretisk Fysik, Uppsala University,
Box 803, S-75108, Uppsala, Sweden}

\address{ $^\dagger$
Institut f\"{u}r Theoretische Physik, ETH-H\"{o}nggerberg,
CH-8093, Z\"{u}rich, Switzerland}

\address{$^\ddagger $
Steklov Mathematical Institute, Fontanka 27, 191011,
St.Petersburg, Russia}

\date{June 1997}
\maketitle

{\tightenlines
\begin{abstract}
In this letter we address the question how interactions
affect the DC conductance of a one-dimensional electron system
not necessarily adequately described by the Luttinger model. 
Using a Laughlin type argument, we show that 
gauge invariance protects the universal value of the 
conductance of $e^2/h$ per channel per spin orientation if 
the system possesses   two conserved charges conjugate to
the chemical potentials of the external reservoirs.

\end{abstract}
}


\narrowtext

The DC conductance of a clean one-dimensional electron system 
described by the  Luttinger model is given by the universal value
of $e^2/h$ per channel and per spin orientation \cite{uc}.
In this letter we address the question whether the  conductance
can deviate from this universal value 
in situations where the Luttinger model may not be
applicable. 

Such a situation may be met, for example, if one considers a 
system with a strong bias, so that the Luttinger fixed point
may no longer describe the state of the system.  
In principle,
one should then take into account an infinite number of irrelevant
operators which were suppressed at the fixed point. One might
expect that this would lead to a nonlinear dependence of the current
on the voltage drop and to a differential conductance depending
on the bias.

Another possibility is the existence of  fixed points
of one-dimensional electron systems different from the Luttinger
models. The latter arise as a result of a perturbative RG analysis
for  short range repulsive potentials. One cannot
exclude the possibility that there exist other conformal theories
describing one-dimensional electrons.

Our analysis is based on a Laughlin type argument \cite{L}
and relate the quantization of the
conductance in a one-dimensional electron system
to gauge invariance. The only conditions  imposed
on the dynamics of the system are the existence of  conserved
charges conjugate to the chemical potentials of the reservoirs
and the absence of impurity backscattering. Under these assumptions,
we show that the conductance is equal
to  $e^2/h$ per channel and per spin orientation.

Recently, experiments were done measuring the conductance in
very pure quantum wires \cite{Yacoby}. At small bias, experimental 
data show that the value of the conductance is somewhat smaller 
than the universal value: $ G \simeq n g(T)2e^2/h,\  n=0,1,2,\dots,$
where $g(T)<1$ is a factor depending on the temperature $T$. 
The factor $g(T)$ can be interpreted as arising 
from a renormalization of the voltage drop and must be explained 
in terms of the physics of the leads and of the junctions between
the wire and the leads.  

To describe the conductance of a one-dimensional system we use
an extension of the Landauer-B\"{u}ttiker approach \cite{LB}
to the case of interacting systems \cite{ACF}. We consider
a one-dimensional system put into  contact with two external
reservoirs with different chemical potentials, $\mu_L$ and $\mu_R$.
We assume that there are two conserved charges, $Q_L$ and $Q_R$,
conjugate to $\mu_L$ and $\mu_R$, which commute with the Hamiltonian
${\cal H}$ of the system and with each other:

\begin{equation}
[{\cal H}, Q_L]=0 \ , \ [{\cal H}, Q_R]=0 \ , \ [Q_L, Q_R]=0.
\end{equation}
Then the thermal state of the system connected to the 
external reservoirs is given
by the density matrix 

\begin{equation} \label{Sig}
\Sigma_{\mu}=\exp(-\beta({\cal H} + \mu_LQ_L +
\mu_R Q_R))
\end{equation}
and its transport properties are
described by {\em equilibrium} statistical mechanics.

A special property of  one-dimensional electric transport is
that the continuity equation for the electric current

\begin{equation}
\p_t \rho + \p_x j=0
\end{equation}
is solved by introducing a scalar operator $\phi$

\begin{equation}
\rho= e \p_x \phi \ , \ j= - e \p_t \phi.
\end{equation}
Here $e$ is the elementary electric charge. This enables one to derive
a formula which is a one-dimensional DC analogue of the Kubo formula.
In the equilibrium state characterized by the chemical potentials
$\mu_L$ and $\mu_R$ the expectation value of the current is
given by

\begin{eqnarray} \label{sr}
\langle j(x) \rangle_{\mu}= - e \langle \p_t \phi(x) \rangle_{\mu} = 
\frac{i}{\hbar} \langle [{\cal H}, \phi(x)] \rangle_{\mu} = \nonumber \\
=\frac{i}{\hbar} ({\rm Tr} \Sigma_{\mu})^{-1} 
{\rm Tr} \left( \Sigma_{\mu} 
[{\cal H}, \phi(x)] \right) = \\
=- \frac{i}{\hbar} \langle [\mu_LQ_L +\mu_RQ_R, \phi(x)]\rangle_{\mu}. 
\nonumber
\end{eqnarray} 
Formula (\ref{sr}) expresses the current in terms of the commutation
relations of the conserved charges with the bosonic field $\phi$
providing  a solution of the continuity equation.

Before applying formula (\ref{sr}) to a general situation, we
consider two simple examples.
We start with a zero-coupling Luttinger model described by the 
Lagrangian

\begin{equation}
{\cal L}= i \hbar \psi_L^* (\p_t- v_F \p_x) \psi_L + 
i \hbar \psi_R^* (\p_t+ v_F \p_x) \psi_R.
\end{equation}
The corresponding Hamiltonian looks as follows

\begin{equation}
{\cal H}_0= i \hbar v_F \int dx
( \psi^*_L \p_x \psi_L - \psi_R^* \p_x \psi_R).
\end{equation}
Standard bosonization formulas,

\begin{eqnarray} \label{bos}
\psi_L^* \psi_L = n_L \ , \ \psi_R^* \psi_R  =  n_R;
 \nonumber \\
i \hbar  \psi^*_L \p_x \psi_L =  \frac{h}{2} n_L^2, \\ 
- i \hbar  \psi_R^* \p_x \psi_R  =  \frac{h}{2} n_R^2 \nonumber
\end{eqnarray}
yield the following expression for  the bosonized Hamiltonian

\begin{equation}
{\cal H}_0= \frac{hv_F}{2}\int dx (n_L^2 + n_R^2).
\end{equation}
Coupling  the system to two external reservoirs which feed
left- and right-moving electrons into the system can be described
by adding to the Hamiltonian an extra term
which includes the 
chemical potentials $\mu_L$ and $\mu_R$ of the reservoirs:

\begin{equation} \label{Hmu}
{\cal H}_{\mu}={\cal H}_0 + \mu_L Q_L + \mu_R Q_R,
\end{equation}
where $Q_L$ and $Q_R$ are conserved particle numbers
of left- and right-moving particles:

\begin{equation}
Q_L= \int dx n_L \ , \ 
Q_R= \int dx n_R.
\end{equation}
Minimizing the expectation value of the Hamiltonian (\ref{Hmu})
gives

\begin{equation}
n_L=-\frac{\mu_L}{hv_F} \ , \ n_R= -\frac{\mu_R}{hv_F}.
\end{equation}
The total current $j=ev_F(n_R-n_L)$ 
is given by the Landauer-B\"{u}ttiker formula

\begin{equation} \label{uc}
j=\frac{e(\mu_L-\mu_R)}{h}=\frac{e^2}{h} V,
\end{equation}
where $V$ is the voltage drop $eV=\mu_L-\mu_R$.

Next, we study the effect of adding an interaction term
described by the following irrelevant operator

\begin{equation}
\D {\cal H} = \e \int dx (\psi^*_L \p_x \psi_L) 
(\psi_R \p_x \psi_R).
\end{equation}
The bosonized interacting Hamiltonian is given by 

\begin{equation}
{\cal H}_{int}={\cal H}_{\mu} + 
\e \pi^2 \int dx n_L^2 n_R^2.
\end{equation}
The expectation value of ${\cal H}_{int}$ achieves its minimum
when

\begin{eqnarray} \label{IIm}
n_L + \frac{2\pi^2 \e}{h v_F} n_L n_R^2 =-\frac{\mu_L}{hv_F}, \nonumber \\
n_R + \frac{2\pi^2 \e}{h v_F} n_R n_L^2 = -\frac{\mu_R}{hv_F}.
\end{eqnarray}

Before computing the current in the interacting system we
should check whether the interaction can possibly change
the definition of the current in terms of $j_L$ and $j_R$.
The current is given by the general formula

\begin{equation}
j(x)= - \left. \frac{\delta {\cal H}}{\delta a(x)}\right|_{a=0} ,
\end{equation}
where $a$ is the spatial component of the vector  potential.
The latter enters the Hamiltonian through the covariant derivatives
$iD_x=i \hbar \p_x + e a$. For instance, the non-interacting Hamiltonian
acquires the form

\begin{equation}
{\cal H}_0(a)= {\cal H}_0 + ev_F\int dx a(n_L-n_R).
\end{equation}
This implies that $j=ev_F(n_L-n_R)$, as we stated before.

By applying the same procedure to the interacting Hamiltonian
we easily arrive at

\begin{eqnarray}
\lefteqn{{\cal H}_{int}(a)={\cal H}_0(a)+} \nonumber \\
& & \ \ \ 
 +\e \pi^2 \int dx (n_L^2 + 2\frac{e}{h} a n_L) 
(n_R^2 -2 \frac{e}{h} a n_R). 
\end{eqnarray}
Therefore, the current gets an extra contribution from the interaction 
term:

\begin{equation}
j_{int}= ev_F(n_R-n_L) - \e \ \frac{2e}{h} n_L n_R (n_R - n_L).
\end{equation}
In the ground state of the interacting 
Hamiltonian (\ref{IIm}), the expectation value of this expression
is equal to $e(\mu_L-\mu_R)/h$.
Thus, the redefinition of the current exactly compensates the extra
contribution caused by the interaction.

Another example of an interacting electronic system is an interacting
Luttinger model which is obtained from the non-interacting one by
adding a marginal perturbation

\begin{equation}
\D {\cal H}_{Lut} = h\e v_F \int dx (\psi^*_L \psi_L) (\psi_R^* \psi_R).
\end{equation}
This perturbation changes some of the bosonization rules (\ref{bos}):

\begin{eqnarray}
i \hbar  \psi^*_L \p_x \psi_L= \frac{h(1+\e)}{2} n_L^2 \\ 
- i \hbar \psi_R^* \p_x \psi_R= \frac{h(1+\e)}{2} n_R^2. \nonumber
\end{eqnarray}
The bosonized Hamiltonian is given by

\begin{eqnarray}
\lefteqn{ \hskip -0.8cm
{\cal H}_{Lut}(a)= hv_F 
\int dx  \left( \frac{1+\e}{2} (n_L^2 + n_R^2) + 
\e n_L n_R \right)+ } \nonumber \\
& & +ev_F \int dx a(n_L-n_R) + \mu_L Q_L +\mu_R Q_R.
\end{eqnarray}
The minimum at $a=0$ is achieved when

\begin{eqnarray}
(1+\e) n_L + \e n_R = - \frac{\mu_L}{hv_F};  \nonumber \\
(1+\e) n_R + \e n_L = - \frac{\mu_R}{hv_F}.
\end{eqnarray}
The current is given by the same universal formula (\ref{uc}).

We have seen that the interaction does not affect the conductance
for two different types of interaction. For the marginal interaction,
the bosonization formulas readjust so as to compensate the
effect of interactions on the current. For irrelevant interactions
considered above, the conductance is not changed because the expression
for the current density in terms of $n_L$ and $n_R$ changes.

Next, we  turn to a proof of the general statement that
formula (\ref{uc}) does not depend on the concrete
dynamics of the one-dimensional electron system, as long
as the reservoirs are coupled to the system via 
conserved particle numbers $Q_L$ and $Q_R$.

Independently of the structure of interactions, the charge density
has the form

\begin{equation}
\rho=e(\psi^*_L \psi_L + \psi^*_R \psi_R).
\end{equation}
We recall that the expression for the electric current
is sensitive to the particular form of interactions.

It is convenient to introduce one-dimensional bosonization formulas
for Fermi fields $\psi_L$ and $\psi_R$:

\begin{eqnarray} \label{pb}
\psi^*_L=e^{2\pi i\phi_L} \ , \ \psi_L=e^{-2\pi i\phi_L};
 \nonumber \\
\psi^*_R=e^{-2\pi i\phi_R} \ , \ \psi_R=e^{2\pi i\phi_R}.
\end{eqnarray}
The bosonic fields $\phi_L$ and $\phi_R$ satisfy the following
commutation relations

\begin{eqnarray} \label{phicom}
 \ [\phi_L(x), \phi_L(y)] =\frac{i}{4\pi}\e (x-y), \nonumber \\
\  [\phi_R(x), \phi_R(y)]  =-\frac{i}{4\pi}\e (x-y), \\ 
\ [\phi_L(x), \phi_R(y)]= \frac{i}{4\pi}, \nonumber 
\end{eqnarray}
where  $\e(x-y)=1, x>y; \e(x-y)=-1, x<y$. The densities of
left- and right-moving particles acquire the form
\begin{equation}
n_L= \p_x\phi_L \ , \ n_R=\p_x \phi_R.
\end{equation}
The conserved charges $Q_L$ and $Q_R$
have the following commutation relations with the bosonic fields:

\begin{equation} \label{Qphi}
[Q_L, \phi_L(x)]=\frac{i}{2\pi} \ , \ [Q_R, \phi_R(y)]=-\frac{i}{2\pi} .
\end{equation}
One can introduce a bosonic field $\phi(x)=\phi_L(x)+\phi_R(x)$
and rewrite the charge density as follows

\begin{equation}
\rho= e(\p_x \phi_L + \p_x \phi_R) =e \p_x \phi.
\end{equation}
We note that all the commutation relations and bosonization
rules listed above depend only on the kinematics of Fermi fields
and are entirely independent of the dynamics of the model we 
consider. The only important assumption which we make is commutativity
of the charges $Q_L$ and $Q_R$ with the Hamiltonian of the
interacting system.

In order to use the definition of the electric current, we should
specify how the vector potential $a$ enters the Hamiltonian.
This is accomplished by a simple substitution

\begin{eqnarray} \label{phia}
\phi_L(x) \rightarrow 
\phi^a_L(x)=\phi_L(x) + \frac{e}{h} \int_{-\infty}^x dy  a(y); \nonumber \\
\phi_R(x) \rightarrow 
\phi^a_R(x)=\phi_R(x) - \frac{e}{h}\int_{-\infty}^x dy  a(y).
\end{eqnarray}
This rule is equivalent to replacing partial derivatives $\p_x$ of fermionic
operators by covariant  derivatives. One can easily check it using (\ref{pb}).
Now we are ready to compute the electric current:

\begin{eqnarray} \label{j}
\lefteqn{ \hskip -0.8cm
j(x)= - \left. \frac{\delta 
{\cal H}(\phi_L^a, \phi_R^a)}{\delta a(x)}\right|_{a=0}=} \nonumber \\
& & =i \frac{e}{\hbar} 
[{\cal H}(\phi_L, \phi_R), \phi_L(x)+\phi_R(x)]= - e \p_t \phi.
\end{eqnarray}
Here we used the Leibniz rule for derivatives and commutators and
the following formulas obtained by combining (\ref{phicom}) and
(\ref{phia})

\begin{eqnarray}
\frac{\delta \phi_L^a(x)}{\delta a(y)}= -i \frac{e}{\hbar}
[\phi_L^a(x), \phi_L(y)+\phi_R(y)]; \nonumber \\
\frac{\delta \phi_R^a(x)}{\delta a(y)} =-i \frac{e}{\hbar}
[\phi_R^a(x), \phi_L(y)+\phi_R(y)].
\end{eqnarray}
The electric current in bosonized form
automatically fulfills the continuity equation:

\begin{equation}
\p_t \rho + \p_x j= e\p_t \p_x \phi - e\p_x \p_t \phi=0.
\end{equation}
Formulas (\ref{Qphi}) and (\ref{j}) can now be used to compute
the conductance from (\ref{sr}):

\begin{eqnarray}
\lefteqn{ \hskip -1cm
 \langle j(x) \rangle_{\mu}
=- i\frac{e}{\hbar} \langle [\mu_LQ_L +\mu_RQ_R, \phi(x)] \rangle_{\mu}=}
\nonumber  \\
&  & \hskip 2cm
 =\frac{e}{h}(\mu_L - \mu_R).
\end{eqnarray}
This confirms the universal conductance formula (\ref{uc})
in the general situation.

This result can be understood by using a simple argument
similar to the one suggested by Laughlin \cite{L} as an
explanation
of the integer Quantum Hall effect. The most general
form of the bosonized Hamiltonian
which commutes with the conserved charges $Q_L$ and $Q_R$
is ${\cal H} = {\cal H}(\rho, \eta, a)$, where $\rho$ is
the charge density and $\eta=(n_L-n_R)$ is the difference
of the particle densities of left- and right-movers. 
Let us for simplicity assume that $\mu_L+\mu_R=0$.

The energy minimum of ${\cal H}_{\mu}$  is achieved if

\begin{equation} \label{min}
\left. \frac{\delta {\cal H}}{\delta \eta(x)} \right|_{\rho}= 
\frac{1}{2} (\mu_R-\mu_L).
\end{equation}
The current in the system is defined by

\begin{equation} \label{cur}
\left. j(x)= - \frac{\delta {\cal H}}{\delta a(x)} \right|_{a=0}.
\end{equation}

In order to relate the left hand side of (\ref{min}) to the
right hand side of (\ref{cur}) one applies  Laughlin's  
argument. More precisely, we assume that our system is put on 
a ring which encircles a magnetic flux $\Phi$. Let us
adiabatically change $\Phi$ by one flux quantum 

\begin{equation}
\Phi \rightarrow \Phi + \frac{h}{e}.
\end{equation}
If the system was in the ground state at the beginning
of the process it will be in a new ground state at the end.
As the change of $\Phi$ by a flux quantum can be compensated 
by a gauge transformation, this will be a new ground state of the {\em same}
Hamiltonian. Formulas (\ref{phia}) imply that in this process
$Q_L= \int dx \p_x \phi_L$ increases by $1$ whereas 
$Q_R= \int dx \p_x \phi_R$ decreases by $1$. Thus, one particle
is transferred from the left-moving channel to the right-moving one.
This implies

\begin{equation} \label{Ae}
\frac{e}{h} da|_{\rho}= \frac{1}{2} d\eta |_{\rho}.
\end{equation}
By combining (\ref{Ae}) with (\ref{min}) and (\ref{cur}) one
obtains the universal result

\begin{equation}
j= - \left. \frac{\delta {\cal H}}{\delta a(x)} \right|_{a=0}
=- \left. \frac{2e}{h} \frac{\delta {\cal H}}{\delta \eta}
 \right|_{\rho}=\frac{e}{h}(\mu_L-\mu_R).
\end{equation}

These considerations are not applicable if the interacting
Hamiltonian includes operators of the type $\cos(2\pi n\phi)$
responsible for backscattering \cite{KF}. Although, such operators
may be irrelevant, they break the symmetry generated by the
charges $Q_L$ and $Q_R$, and hence formula (\ref{Sig}) for 
the thermal state is no longer applicable.

\widetext

\end{document}